\documentstyle[12pt]{article}
\begin{document}

\title{Nonlocal Phenomena from Noncommutative Pre-Planckian
Regime} 

\author{%
Michael Heller\thanks{Correspondence address: ul.
Powsta\'nc\'ow Warszawy 13/94, 33-110 Tarn\'ow, Poland, e-mail:
mheller@wsd.tarnow.pl} \\
Vatican Observatory, \\ V-00120 Vatican City State \\
\and
Wies{\l}aw Sasin \\
Institute of Mathematics, \\ Warsaw University of Technology, \\
Plac Politechniki 1, 00-661 Warsaw, Poland}

\maketitle

\begin{abstract}
A model unifying general relativity with quantum mechanics is
further developed. It is based on a noncommutative geometry
which supposedly modelled the universe in its pre-Planckian
epoch. The geometry is totally nonlocal with no time and no
space in their usual meaning. They emerge only in the transition
process from the noncommutative epoch to the standard space-time
physics.  Observational aspects of this model are discussed.
It is shown that various nonlocal phenomena can be explained as
remnants of the primordial noncommutative epoch. In particular,
we explain the Einstein-Podolsky-Rosen experiment, the horizon
problem in cosmology, and the appearance of singularities in
general relativity.
\end{abstract}

\newpage


\section{Introduction}
There are many hints coming from several independent research
programs that in the quantum gravity regime the standard
concepts of space and time break down.  Following this
suggestion we propose a scheme for unification of general
relativity and quantum mechanics based on a radically nonlocal
geometry in which such structures as space point, time instant,
and their neighborhoods, cannot be even defined.  Namely, we
assume that it is a noncommutative geometry (see \cite{Landi}
for an introductory course and \cite{Connes} for a comprehensive
monograph) that correctly models the universe below the Planck
threshold, and only during the ``phase transition'' from a
noncommutative geometry to the standard (commutative) geometry,
space, time and other local structures emerge. In \cite{HSL} we
have proposed a concrete model implementing the above ideology.
Although based on a sound mathematical basis it should be
treated as a toy model (because of some both conceptual and
computational simplifications).  It turns out that some nonlocal
phenomena, known from quantum mechanics and cosmology, such as
the Einstein-Podolsky-Rosen (EPR) type of experiments, the horizon
problem, and the appearance of classical singularities, can be
explained as remnants (or ``shadows'') of the primordial global
physics. To discuss these phenomena in terms of a noncommutative
geometry is the goal of the present paper. Although the
analysis is based on our model presented in \cite{HSL}, it is
independent of the particulars of this model. In section 2, we
give a brief review of our model, and in sections 3 through 5 we
discuss, in turn, the EPR experiment, the horizon problem, and
the classical singularity problem.

\section{An Overview of the Model}
Let $\pi_M:E\rightarrow M$ be a fiber bundle of orthonormal
frames over a space-time manifold $M$, and $\Gamma$ a group.  In
the following, for concreteness, we shall assume that
$\Gamma =$SO(3,1), but the correct choice of $\Gamma$ should be
based on physical grounds and is left for the future
developments of the model.  The Cartesian product $
G=$$E\times\Gamma$ has the structure of the groupoid (unlike in
a group its elements can be composed only if they belong to the
same fiber).  We define the involutive algebra ${\cal A}$ of
smooth, complex valued, compactly supported functions on $G$
with the convolution
\[(a*b)(\gamma ) :=\int_{G_p}a(\gamma_1)b(\gamma_2)\]
as multiplication, where $a,b\in {\cal A},\,\gamma
=\gamma_1\circ\gamma_2,\gamma ,\gamma_1,\gamma_ 2\in G_p$, $G_p$
is the fiber in $G$ over $p\in E$, and the integral is taken
with respect to the (left) invariant Haar measure. The
involution in ${\cal A}$ is defined in the following way:
$a^{*}(\gamma ):=\overline {a (\gamma_{}^{-1})}$.  The 
noncommutative algebra ${\cal A}$ is an analogue of the algebra
of smooth functions on a manifold and serves us to construct a
noncommutative geometry of our model.

Let $V$ be a subset of derivations of ${\cal A}.$ A derivation
$v \in V$ is defined to be a linear mapping $v:{\cal A}
\rightarrow {\cal A}$ satisfying the Leibniz rule.  
It can be interpreted as a counterpart of a vector field.  We
assume that $V$ is of the form $V=V_E\oplus V_{\Gamma}$ where $
V_E$ are derivations ``parallel to $E$'' and $V_{\Gamma}$
derivations ``parallel to $\Gamma$''. Basing on $V$ one can
define all fundamental concepts of differential geometry, such
as: linear connection, curvature, Ricci tensor and,
consequently, one can write a noncommutative generalization of
Einstein's equation.  In the natural way, this geometry splits
into the part ``parallel to $ E$'' and the part ``parallel to
$\Gamma$'' (for details see \cite{Towards,HSL} and the
bibliography cited therein).

The special care should be given to the metric problem.  It
has been showed by Madore and Mourad \cite{MadMour} that for a
broad class of derivation based differential geometries the
metric is essentially unique.  It turns out that in our case
this applies to the $\Gamma$-parallel part of the metric
$g_{\Gamma}$, whereas the $ E$-parallel part of the metric $g_E$
behaves in the standard way (it is essentially lifting of the
metric from space-time $M$).  Einstein's equation surprisingly
well cooperates with this fact.  It is an operator equation
which should be solved for derivations (not for a metric!).
However, since all derivations belonging to $V_E$ solve the
$E$-parallel part of Einstein's equation this part of the
equation becomes an equation for metric (for details see
\cite{Towards}).

Let $G_p$ and $G_q$ be two fibres of $G$ over $ p,q\in E$,
respectively.  This fibres are said to be equivalent if there is
$ g\in\Gamma$ such that $p=qg$.  Let ${\cal A}_{proj}$ be the
subalgebra of ${\cal A}$ consisting of all functions which are
constant on the equivalence classes of this relation.  ${\cal
A}_{proj}$ is a subset of the center of the algebra ${\cal A}$
and thus a commutative algebra (in this case, convolution
becomes the usual multiplication).  It can be shown that the
algebra ${\cal A}_{proj}$ is isomorphic with the algebra $
C^{\infty}(M)$ of smooth functions on the space-time manifold
$M$.  In this way, we recover the standard space-time geometry
and the standard theory of general relativity (in the Geroch
formulation
\cite{Geroch}).

Surprisingly, the $\Gamma$-part of the above geometry leads to
the quantum mechanical approximation.  To make the contact with
the usual Hilbert space formulation of quantum mechanics let us
define the representation $\pi_q\! :\, {\cal A}\rightarrow {\cal
B}({\cal H} )$ of the algebra ${\cal A}$ in the Hilbert space $
$$ {\cal H}=$$ $$L^2(G_q),q\in E$, where ${\cal B}({\cal H})$
denotes the algebra of bounded operators on $ {\cal H}$, by the
following formula
\begin{equation}(\pi_q(a)\psi )(\gamma )=\int_{G_q}
a(\gamma_1)\psi (\gamma_1^{-1}\gamma
),\label{repr}\end{equation} 
$a\in {\cal A},\,\psi\in {\cal H}$.  It is worthwhile to notice
that the completion of $ {\cal A}$ with respect to the norm
$\parallel a\parallel = {\rm sup}_{q\in
E}\parallel\pi_q\parallel$ is a $C^{ *}$-algebra. It turns out
that if we restrict our analysis to the ``wave functions'' $\psi
$ which are $\Gamma$-invariant, i. e., which are constant on the
equivalence classes of fibres of $ G$, we essentially obtain
quantum mechanics in the Heisenberg picture (see
\cite{Towards,HSL}).

We can assume that to every derivation $v\in V$ there
corresponds an internal derivation in the image $\pi_q({\cal A}
)$ given by
\begin{equation}i\hbar\pi_q(v(a))=[F_v,\pi_q(
a)], \label{dyn}\end{equation}
for every $a \in {\cal A}$, where $F_v$ is a self-adjoint
operator (satisfying certain ``technical'' conditions); the
factor $i\hbar$ is added for the future convenience.  If
$\psi\in L^2(G_q)$ is $\Gamma$-invariant, $v$ the usual
differentiation (with respect to time in a certain coordinate
system), and $F_v$ the Hamiltonian of the system then eq.
(\ref{dyn}) becomes the usual Schr\"odinger equation in the
Heisenberg picture of quantum mechanics.

\section{The EPR Experiment}
In the following we shall focus on those nonlocal effects of the
noncommutative pre-Planck epoch that can survive the transition
to the weak gravity approximation of our model. First, we shall
be interested in those effects which make themselves manifest in
the quantum mechanical domain. This means that we should
consider the subalgebra ${\cal A}_{\Gamma}$ consisting of those
functions on $ G$ which are lifting of smooth compactly
supported functions on the group $\Gamma$, i.  e.,  ${\cal
A}_{\Gamma}:= \{f\circ pr_{\Gamma}:f\in C^{\infty}_c(\Gamma
,{\bf C})\}\subset {\cal A}$, where $pr_{\Gamma}
:G\rightarrow\Gamma$ is the canonical projection. Let $a\in
{\cal A}_{\Gamma}$ and $ p,q\in E$, $p\neq q$, and let us
consider the following representations of ${\cal A}_{
\Gamma}$ 
\begin{equation}\pi_p(a)(\psi_p)=a_p*\psi_p\label{reprp}\end{equation}
\begin{equation}\pi_q(a)(\psi_q)=a_q*\psi_q\label{reprq}\end{equation}
where $\psi_p\in L^2(G_p),\,\psi_q\in L^2(G_q )$.  $G_p$ and
$G_q$ are isomorphic; therefore, $\psi_p$ and $\psi_q$ can be
chosen to be isomorphic as well.  This, in turn, implies that
$\pi_p(a)$ and $\pi_q(a)$ are also isomorphic.  Therefore, if $
a\in {\cal A}_{\Gamma}$ then its image under the representation
$\pi_ p$ does not depend on the choice of the fibre $G_p$ (up to
isomorphism).  Since $ p\in E$ projects down to $\pi_M(p)\in M$,
the above can be rephrased by saying that all points of
space-time $M$ ``know'' what happens in the entire fibre $
G_g,\,g\in\Gamma$.  The conclusion is that the observational
consequences of the proposed model should be looked for among
correlations between distant events in space-time rather than
among local phenomena (for details see \cite{EPR}).  In the
following we shall consider some examples of such correlations.

As the first example we show that the famous
Einstein-Podolsky-Rosen effect, experimentally verified by Alain
Aspect and others, can be deduced from our model (see
\cite{EPR}).  Let us assume that $\Gamma_0=SU(2)$ is a subgroup
of $\Gamma$, and let us choose two linearly independent
functions on $\Gamma_ 0$ which span the linear space ${\bf
C}^2${\bf .}  Let further $\hat {S}_z=\pi_ p(s)|_{{\bf
C}^2},\,s\in {\cal A}$, be the z-component of the usual spin
operator. By using representation (\ref{repr}) we can write the
corresponding eigenvalue equation in the following form
\[\int_{\Gamma_0}s_p(\gamma_1)\psi (\gamma_1^{
-1}\gamma )=\pm\frac {\hbar}2\psi .\]
Remembering that $s_p={\rm const} $, as one of the solutions of
this equation we obtain $\psi ={\bf 1}_{\Gamma_0}$, and
consequently
\[\frac {\hbar}2=\pm \int_{\Gamma_0} s_p(\gamma_1).\]
Hence, $(s_p)_1=+(\hbar /2)$(${\rm v}{\rm o}{\rm l}
\Gamma_0)^{-1}$, $(s_p)_2=-(\hbar /2)({\rm v}
{\rm o}{\rm l}\Gamma_0)^{-1}$. The corresponding eigenvalue
equations are
\[\pi_p((s_p)_1)\psi =+\frac {\hbar}2\psi\;\;
{\rm f}{\rm o}{\rm r}\;\psi\in {\bf C}^{+},\]
\[\pi_p((s_p)_2)\psi =-\frac {\hbar}2\psi\;\;
{\rm f}{\rm o}{\rm r}\;\psi\in {\bf C}^{-}\]
where ${\bf C}^{+}:={\bf C}\times \{0\},$ and $ $$
{\bf C}^{-}:=\{0\}\times {\bf C}$.

Let us consider an observer $A$ who is situated at a point
$x_A=\pi_M(p)$, $p\in E,$ in space-time $ M$, and an observer
$B$ who is situated at a point $x_B=\pi_M(q),\,q\in E$, in M
distant from the point $x_A$. The observer $A$ measures the
z-spin component of the one of the electrons in the EPR
experiment, i. e., he acts with the operator $
\hat {S}_z\otimes {\bf 1}|_{{\bf C}^2}$ on the state
vector $\xi =\frac 1{\sqrt {2}}(\psi\otimes\phi -\phi\otimes\psi
)$ where $\psi\in {\bf C}^{+}$ and $\phi\in {\bf C}^{-}$. Let us
assume that the result of the measurement is $+(\hbar /2)$.
Therefore, $\xi$ collapses to the state vector $\xi_ 0=\frac
{\hbar}{\sqrt {2}}(\psi\otimes\phi )$. However, this vector is
the same (up to isomorphism) for all fibres $ G_r,\,r\in E$ (let
us notice that the analogues of formulae (\ref{reprp}) and
(\ref{reprq}) are also valid for tensor products). In
particular, $\xi_0$ is the same for the fibres $ G_p$ and $G_q$.
Now, if the observer $B$, situated at $x_B$, measures the z-spin
component of the second of the electrons, i. e., if he acts with
the operator ${\bf 1}|_{{\bf C}^ 2}\otimes\hat {S}_z$ on the
vector $\xi_0$, the only possible result of the measurement
could be $-(\hbar /2)$.

In this approach there is no information transfer between $ A$
and $B$, but rather the process of measurement somehow relates
to the fundamental level which is atemporal and aspatial.

\section{The Horizon Problem}
Another example which naturally comes to the mind is the horizon
paradox in relativistic cosmology. If the present universe
evolved from the pre-Planckian nonlocal stage it is
straightforward to expect that various parameters determining
its structure at various places are correlated even if these
places were never, after the Planck epoch, in any causal
contact. In particular, this refers to the ``uniformity'' of the
universe. Let $\rho\in {\cal A}$ be an observable corresponding
to this ``uniformity''; for instance, let $\rho$ be related to
the density of the universe. To be an observable $
\rho$ must be an Hermitian element of ${\cal A}$, and to leave
traces in space-time $\rho$ must be an element of the subalgebra
${\cal A}_{proj}$. The eigenvalue equation of the observable $
\rho$ is
\[\int_{G_q}\rho (\gamma_1)\psi (\gamma_1^{-1}
\gamma )=r_q\psi (\gamma ).\]
Since, in this case, $\psi$ is $\Gamma$-invariant it is easy to
calculate that
\begin{equation}r_q=\int_{G_q}\rho (\gamma_1)
=\rho (\gamma_1){\rm v}{\rm o}{\rm l}\Gamma
.\label{eigv}\end{equation} 

Let us define the ``total phase space'' of our system:
$L^2(G):=\bigoplus_{q\in E}L^2(G_q)$ with the operator $
\pi (\rho ):=(\pi_q(\rho ))_{q\in E}$ acting on it.  
Now, eq.  (\ref{eigv}) assumes the form $r(x) :=r_q$, where
$\pi_M(q)=x\in M$; it defines the function $r:M\rightarrow
{\bf R}$ on $ M$.  Consequently, if the measurement
corresponding to $\rho$ is performed at a point $ x\in M$ its
result $r(x)$ is correlated with the result $r(y)$ of another
measurement of the same quantity at another point $y\in M$ even
if these two points are separated by the horizon (in the sense
that both $ r(x)$ and $r(y)$ are the values of the same
function). If one postulates that $r(x)$ is a constant function,
it must be proportional to ${\rm v}{\rm o}{\rm l} \Gamma$. In
fact, to solve the horizon paradox it is enough to postulate
that $r(x)$ is a sufficiently slowly varying function of $x$
(for the sake of simplicity, we do not take into account effects
of possible ``fluctuations'').

\section{The Classical Singularity Problem}
Another ``global problem'' which can be explained by the
proposed model is the classical singularity problem in general
relativity.  For a long time it was known that singularities of
stronger types (e.  g., curvature singularities) have truly
global properties.  For instance, the existence of such global
features of space-time as the appearance of Cauchy or particle
horizons depends on the structure of singularities.  They are
usually defined as ideal points of space-time or their singular
boundaries, and consequently it is more reasonable to speak
about singular space-times rather than about singularities in
space-time (see \cite{HawEllis}).  These aspects of the
singularity problem suggest that they could somehow be related
to the nonlocal physics of the pre-Planckian era.  This indeed
turns out to be the case.

Let us construct the generalized fibre bundle of orthonormal
frames on a space-time $ M$ with its singular boundary $\partial
M$ (for instance Schmidt's b-boundary), $\bar {M}=M\cup\partial
M$.  Let $E$ be the total space of this generalized fibre
bundle.  The fibres of $E$ over $\partial M$ are degenerate (in
the case of the closed Friedman universe and Schwarzschild's
solution with their b-boundaries these fibres degenerate to the
single points).  However, it can be shown that if we construct
the groupoid $G$$=E\times\Gamma$, the fibres of $ G$ over
degenerate fibres of $E$ are regular (for details see
\cite{HSJMP}).  This procedure truly deserves the name
``desingularization process''. Now, we can define the algebra
${\cal A}$ of smooth, complex valued, compactly supported
functions on $ G$, and proceed exactly in the same way as above.
In the noncommutative regime there are no points, but there are
states of the system (represented by the states on the algebra
${\cal A}$, i. e, by positive, suitably normed functionals on
${\cal A}$), however with no possibility to distinguish between
singular and non-singular states \cite{HSJMP,HSGRG}. In these
circumstances it is meaningless to speak about singularities in
any sense.

If we reverse this construction, starting from the
non-commutative geometry of the groupoid $G$ and going down to
the space-time manifold, we can clearly see that singularities
arise in the process of taking the double quotient by the action
of the group $\Gamma$: first to obtain $E=G/\Gamma$, and then to
obtain the space-time with singularities $\bar {M}=E/\Gamma$
(this is studied in \cite{HSGRG})

The question: will the future theory of quantum gravity remove
singularities from our picture of the universe? is usually
presupposed to admit two answers:  the answer ``yes'' has become
a kind of common wisdom; the answer ``no'' is adopted only by a
few.  The above results open the third possibility.  On the
fundamental level (below the Planck threshold) it is meaningless
even to ask about singularities.  The noncommutative geometry
shaping this level is totally nonlocal:  there is no space, no
time (in their usual sense) and no distinction between singular
and non-singular states.  In spite of this, the true (albeit
generalized) dynamics is possible (see \cite{HSPhLet}).  It is
only in the transition process from the noncommutative regime to
the commutative geometry (the sort of the first phase
transition), when the space-time emerges and its singular
boundaries are produced.  From the ``point of view'' of the
fundamental level there are no singularities; from our point of
view, who are macroscopic observers, there was the singularity
in the beginning of our universe, and possibly there will be one
at its end.

\end{document}